# The effect of the spin and orbital parts of the Poynting vector on light localization in solid - core micro – structured optical fibers

**G. K. A**LAGASHEV,[1] **S. S. S**TAFEEV,[1,2] **V. V. K**OTLYAR[2] AND **A. D. P**RYAMIKOV[1*]

[1]*Prokhorov General Physics Institute of the Russian Academy of Sciences, 38 Vavilov street, Moscow, 119333, Russia*
[2]*Image Processing Systems Institute of the Russian Academy of Sciences – Branch of the FSRC "Crystallography and Photonics"RAS, 151 Molodogvardejskaya street, Samara, 443001, Russia*
*\*pryamikov@mail.ru*

**Abstract:** The optical properties of solid – core micro – structured optical fibers (SC MOFs) have been studied for a long time. The process of energy outflow of the core modes has always been associated with the process of constructive interference of the core modes fields under reflection from the photonic crystal cladding. In this paper, we want to offer a new look at the light localization in the core of SC MOFs related to the behavior of spin and orbital parts of the Poynting vector of these core modes and singularities arising in it.

## 1. Introduction

The optical properties of SC MOFs have been extensively researched for years. Two scientific areas that are currently significantly grown are fiber-optics communications, which investigates the transmission of information through optical fibers, and singular optics, which researches the propagation of beams with a phase singularities [1].

At the intersection of these two areas there are studies on the propagation of beams with orbital angular momentum in optical fibers. Due to the independent propagation of the beams with distinct topological charges [2], these beams can be used to compress a signal transmitted over a fiber in optics communications [3,4]. To explain the propagation of the OAM-modes different types of fibers were previously introduced: such as a ring-shaped fiber [5,6], ring-core photonic crystal fiber [3], a multicore fiber [7], negative curvature ring-core fiber [8], hollow-core photonic bandgap fiber [9], an inverse-parabolic graded index fiber [10], etc. It should be noted that it is not only the propagation of classical phase optical vortices in fibers that has been investigated, but also polarization vortices (or high-order cylindrical vector beams) [7,10].

The energy flow in the cross section (transverse energy flow) of the optical vortex is rotated, with transverse energy flows being currently actively researched for free-space propagated beams [11–15]. Previously it was shown [16,17] that the energy flow (or the Poynting vector) is equal to the sum of two flows: the orbital energy flow and the spin energy flow. This representation was introduced for the first time by Bekshaev and Soskin in [18] for paraxial fields. Later this approach was adapted to non-paraxial vector fields by Berry [16]. The transformation of one type of flow to another one is possible due to the spin-orbit interaction [19,20].

In this paper, we examine the behavior of the spin and orbital parts of the Poynting vector of the fundamental leaky core mode for holey fibers (HFs) and all solid band gap fibers (ASBGFs) [21] and demonstrate their influence on the degree of light localization in these fibers. We also demonstrate that the waveguide losses in such a simple waveguide as an all solid dielectric pipe with the core refractive index lower than the surrounding medium are determined by the radial projection of the orbital part of the Poynting vector of the fundamental core mode.

In our earlier paper [22], it was shown that for ASBGFs there are sets of geometric parameters such as a pitch value and a cladding rod diameter that allow to obtain losses of the fundamental core mode two or three orders of magnitude lower in the narrow spectral regions than in neighboring transmission bands. In the case of HFs, this phenomenon has not been observed, with losses always diminishing monotonically with a decreasing wavelength. The key point in understanding the resonant loss reduction in ASBGFs lies in explaining the deviation of the streamlines of the orbital part of the transverse component of the core mode Poynting vector from the radial direction. In the case of resonant loss reduction, the orbital part of the Poynting vector of the fundamental core mode has singularities in the cross – section of the fiber and vortices are observed in the cladding rods. At the same time, vortex motions of the orbital part of the transverse component of the Poynting vector of the fundamental core mode are not observed in HFs with the same geometrical parameters as for ASBGFs. Depending on the symmetry of the cladding elements arrangements in the cladding of ASBGFs and the number of the cladding rod layers, several narrow transmission bands with very low losses can be observed [22].

The paper has 4 sections: in Section 2 we describe orbital and spin parts of the Poynting vector of the fundamental core mode of the solid core dielectric pipe and demonstrate their influence on the waveguide losses; in Section 3 the orbital and spin parts of the Poynting vector of the core modes of HFs and ASBGFs are calculated and the effect of their vortex motions on the fiber losses is demonstrated; Section 4 is Conclusions.

## 2. Orbital and spin parts of the transverse Poynting vector component of the fundamental core mode of an all-solid dielectric pipe. Their effect on waveguide losses

In order to understand the general principles of behavior of the orbital and spin parts of the transverse component of the Poynting vector of the core modes of leaky waveguides, let us consider the simplest leaky waveguide, namely, an all - solid dielectric pipe. Let the refractive index of the pipe core be $n_1 < n_2$, where $n_2$ is a refractive index of the surrounding medium. In this case, the transverse component of the electric and magnetic fields of the core modes of $E_r, H_r$ and $E_\varphi, H_\varphi$ can be expressed in terms of the axial component of these fields of $E_z$ and $H_z$. We will consider the transverse component of the Poynting vector of the fundamental core mode, since it is responsible for the loss level in SC MOFs. The transverse component of the Poynting vector can be obtained from the following expressions for the total Poynting vector [16, 18]:

$$\vec{P} = \frac{c^2}{2\omega}\varepsilon_0 \operatorname{Im}\left[\vec{E}^* \times \left(\vec{\nabla} \times \vec{E}\right)\right], \qquad [\text{Eq. (1)}]$$

where $\vec{P} = \vec{P_t} + P_z\vec{z}$ and $\vec{E} = \vec{E_t} + E_z\vec{z}$ is a total electric field of the core mode. The Poynting vector Eq.(1) can, in turn, be decomposed into orbital and spin parts that have transverse components [16]:

$$\vec{P}^{orb} = \frac{c^2}{2\omega}\varepsilon_0 \operatorname{Im}\left[\vec{E}^*\left(\vec{\nabla}\right)\vec{E}\right], \qquad [\text{Eq. (2)}]$$

$$\vec{P}^{spin} = \frac{c^2}{4\omega}\varepsilon_0 \vec{\nabla} \times \text{Im}\left(\vec{E}^* \times \vec{E}\right).$$

where $\vec{P}^{orb} = \vec{P}_t^{orb} + P_z^{orb}\vec{z}$ and $\vec{P}^{spin} = \vec{P}_t^{spin} + P_z^{spin}\vec{z}$.

If we express the transverse components of the electric field of an arbitrary core mode of the all - solid leaky pipe in Cartesian coordinates via the components of this field in cylindrical coordinates, it is easy to show that for linear polarization and circular polarization of the core modes, the spin part of the Poynting vector is not equal to zero, contributing to the transverse and longitudinal components of the Poynting vector. The transverse components of the electric fields of the core mode can be expressed in cylindrical coordinates as:

$$E_x = E_r \cos\varphi - E_\varphi \sin\varphi, \qquad [\text{Eq. (3)}]$$
$$E_y = E_r \sin\varphi + E_\varphi \cos\varphi,$$

where $\varphi$ is an azimuthal angle in the cylindrical coordinate system. The all - solid dielectric pipe is a leaky waveguide, with all the components of the electric field of the core modes being complex functions. Then, we will characterize the polarization state of the leaky core modes using the ratio:

$$\frac{E_x}{E_y} = S, \qquad [\text{Eq. (4)}]$$

where $S$ is a complex or real number, which can change near the waveguide boundary. Here it is assumed that $E_x, E_y \gg E_z$. Using Eq.(3) we can obtain a ratio for the radial and azimuthal components of the electric fields of the core modes of an all - solid dielectric pipe:

$$\frac{E_r}{E_\varphi} = \frac{(\sin\varphi + S\cos\varphi)}{(\cos\varphi - S\sin\varphi)} = F(\varphi), \qquad [\text{Eq. (5)}]$$

where $F(\varphi)$ is a complex function.

Substituting Eq.(5) in Eq.(2) we can obtain three components of $\vec{P}^{spin}$ taking into account the phase factor $e^{i(\beta z - \omega t)}$ for all components of the core mode electric field. The propagation constant is $\beta = \beta^{\text{Re}} + i\beta^{\text{Im}}$ or $\beta = \left(\text{Re}(n_{eff}) + i\,\text{Im}(n_{eff})\right)\omega/c$, where $n_{eff}$ is an effective mode index. For linear polarization of the fundamental core mode $S$ is real and function $F(\varphi)$ is also real. Then, the projections of the spin part of the Poynting vector of the core modes are determined as follows:

$$P_r^{spin} = \frac{c^2}{2\omega}\varepsilon_0 \beta^{\text{Im}} A, \qquad [\text{Eq. (6)}]$$

$$P_\varphi^{spin} = -\frac{c^2}{2\omega}\varepsilon_0 \beta^{\text{Im}} B,$$

$$P_z^{spin} = \frac{c^2}{4\omega}\varepsilon_0 \frac{1}{r}\left[\frac{\partial[rA]}{\partial r} - \frac{\partial B}{\partial \varphi}\right],$$

where $A = F(\varphi)\text{Im}\left(E_\varphi E_z^* - E_\varphi^* E_z\right)$ and $B = \text{Im}\left(E_\varphi^* E_z - E_\varphi E_z^*\right)$.

Eq.(6) show that transverse components of the spin part of the Poynting vector of the core modes are much smaller than the axial component because $\beta^{\text{Im}}/k_0 \ll 1$ in the leaky waveguides, where $k_0 \approx \omega/c$. It means that spin part of the transverse component of the Poynting vector should have a little effect on the loss level of the core mode of an all - solid dielectric pipe in the case of linear polarization.

As for the orbital part of the transverse component of the Poynting vector of the core modes of a dielectric pipe, it is easier to express it immediately in cylindrical coordinates from Eq.(2):

$$P_r^{orb} = \frac{c^2}{2\omega}\varepsilon_0 \operatorname{Im}\left[E_r^* \frac{\partial E_r}{\partial r} + E_\varphi^* \frac{\partial E_\varphi}{\partial r} + E_z^* \frac{\partial E_z}{\partial r}\right],$$

$$P_\varphi^{orb} = \frac{c^2}{2\omega}\varepsilon_0 \frac{1}{r}\operatorname{Im}\left[E_r^* \frac{\partial E_r}{\partial \varphi} + E_\varphi^* \frac{\partial E_\varphi}{\partial \varphi} + E_z^* \frac{\partial E_z}{\partial \varphi}\right], \quad \text{[Eq. (7)]}$$

$$P_z^{orb} = \frac{c^2}{2\omega}\varepsilon_0 \operatorname{Im}\left[E_r^* \frac{\partial E_r}{\partial z} + E_\varphi^* \frac{\partial E_\varphi}{\partial z} + E_z^* \frac{\partial E_z}{\partial z}\right].$$

Since, the electric field components of the fundamental core mode of the dielectric pipe have such dependencies on the azimuthal angle as $\cos\varphi$ or $\sin\varphi$, the $\varphi$ – component of the orbital part of the Poynting vector should be equal to zero. Thus, the orbital part of the Poynting vector of the core modes is determined only by the radial and axial projections in the case of linear polarization.

For circular polarization of the core modes, the function $F(\varphi)$ is equal to $\pm i$ Eq.(5). Equations for transverse projections of the spin part of the Poynting vector will have the following form:

$$P_r^{spin} = \frac{c^2}{4\omega}\varepsilon_0 \left(\frac{1}{r}\frac{\partial C}{\partial \varphi} + 2\beta^{\operatorname{Im}} A\right), \quad \text{[Eq. (8)]}$$

$$P_\varphi^{spin} = \frac{c^2}{4\omega}\varepsilon_0 \left(-2\beta^{\operatorname{Im}} B - \frac{\partial C}{\partial r}\right),$$

where $C = \mp 2i|E_\varphi|^2$. For circular polarization of the core modes, Eq.(8) indicate that the value of the spin part of the transverse component of the Poynting vector is no longer determined by the small term of the order of $\beta^{\operatorname{Im}}/k_0 \ll 1$, although the radial part in Eq.(8), as $r \to \infty$, gives the same answer as in the case of linear polarization Eq.(6). This means that the contribution to the waveguide losses from the spin parts of the Poynting vector is the same for both types of the polarization.

Let us analyze the behavior of spin and orbital parts of the transverse component of the Poynting vector of the fundamental core mode of a dielectric pipe and study their effect on the outflow of the fundamental core mode energy (waveguide losses). The numerical calculations are carried out for linear and circular polarization of the fundamental core mode. To calculate the spin and orbital parts of the Poynting vector of the fundamental core mode we used Comsol Multiphysics and the vector BPM method realized in the BeamPROP.

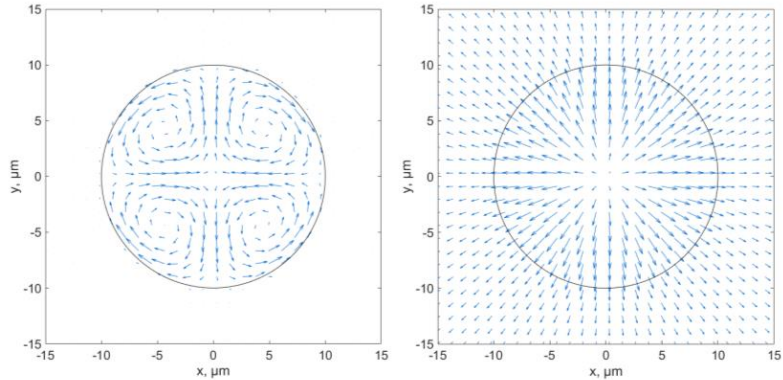

Fig. 1. Spin (left) and orbital (right) parts of the transverse component of the Poynting vector of the fundamental core mode of a dielectric pipe in the case of linear polarization. The refractive index of the pipe core is 1.45 and the refractive index of the surrounding medium is 1.5. The pipe core diameter is 20 μm and the wavelength is 1 μm.

Both transverse components of the Poynting vector of the fundamental core mode for linear polarization are shown in Fig. 1. As was mentioned above, it is only in the presence of the core mode energy outflow it is possible to obtain a non – zero spin part of the Poynting vector Eq.(6). As can be seen from Fig. 1(left), the spin part of the transverse component of the Poynting vector is not equal to zero and has four singularities in the cross – section of the pipe. The orbital part of the transverse component of the Poynting vector of the fundamental core mode has a nonzero radial component Eq.(7) and therefore the energy that the mode loses when propagating through the pipe is mainly related with the orbital part (Fig. 1(right)). For the orbital part of the transverse component of the Poynting vector there is only one singularity on the pipe axis.

In the case of circular polarization of the core modes, the behavior of the spin part is mainly determined by the derivatives of the intensity of the azimuthal electric field component (Fig.2 (left)). In this case, there is one singularity in the cross – section which is also found at the origin. The orbital part of the transverse component of the Poynting vector of the fundamental core mode has the same behavior as in the case of linear polarization with one singularity at the origin (Fig. 2(right)).

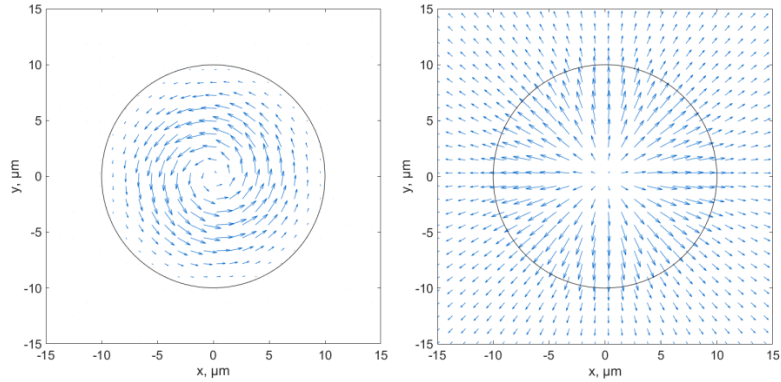

Fig. 2. Spin (left) and orbital (right) parts of the transverse component of the Poynting vector of the fundamental core mode of a dielectric pipe in the case of circular polarization. The refractive index of the pipe core is 1.45 and the refractive index of the surrounding medium is 1.5. The pipe core diameter is 20 μm and the wavelength is 1 μm.

Comparing $r$ - and $\varphi$ - projections of spin and orbital parts of the Poynting vector Eq.(6-8), can be concluded that the outflow of the core modes energy in the dielectric pipe and correspondingly their losses is mainly determined by the $r$ – projection of the orbital part. The axial energy flow of the core modes is determined by the z – projections of both spin and orbital parts of the Poynting vector.

### 3. Spin and orbital parts of the Poynting vector of the core mode for ASBGFs and HFs. Low loss regimes.

ASBGFs have one unusual feature associated with the existence of such sets of geometric parameters (cladding rod radius and pitch Λ) at a given refractive index contrast between the core and the cladding rods that allow to obtain very low loss regimes even with a single layer of the cladding rods (Fig. 3(left)). Let us consider an ASBGF with the same refractive index contrast between the core and the cladding rods as in our work [22], where $n_{core}$ = 1.45 and $n_{rod}$ = 1.5, and demonstrate the loss dependence on values of the cladding rod radius and the pitch. The waveguide losses are determined by $\text{Im}(n_{eff})$ and its dependence on the geometric parameters of the fiber is shown in Fig. 4 at a wavelength of 1 μm.

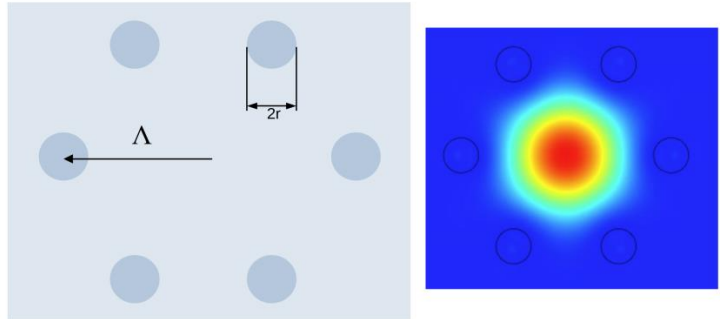

Fig. 3. (Left) Cross – section of an ASBGF and an HF with one row of the cladding holes and rods in the cladding; (right) fundamental core mode of an ASBGF.

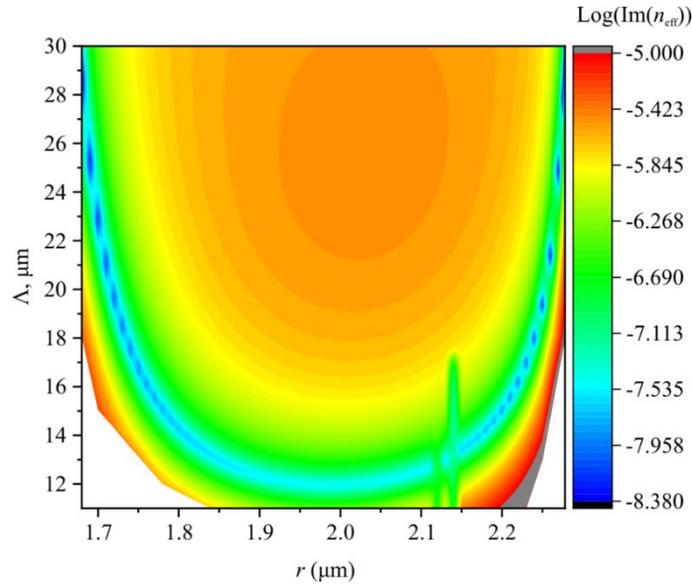

Fig. 4. Dependence of waveguide losses of the fundamental core mode of an ASBGF with one row of the cladding rods on values of the cladding rod radius and the pitch at the refractive index contrast 0.05; the refractive index of the glass matrix is 1.45. The calculation was carried out at a wavelength of 1 μm for circular polarization. The loss behavior near the values of the cladding rod radius of 2.15 μm is related to anti – crossing between the fundamental core mode and the cladding rod modes.

Fig. 4 shows a narrow range of values of the cladding rod radius and the pitch at which the losses of the fiber fundamental core mode fall by orders of magnitude compared to neighboring loss values. This loss behavior dependence on the geometrical parameters of the fiber cannot be explained by the reflection of the fundamental mode radiation from the cladding rods. To demonstrate this, a similar waveguide loss dependence with the same geometric parameters was calculated for a HF with a single layer of the cladding holes. The results of the calculation are shown in Fig .5.

It can be seen from Fig. 5 that HFs with the same geometrical parameters do not have the same features in the waveguide loss dependence as the ASBGF with the waveguide loss behavior shown in Fig. 4. In this case, waveguide losses decrease monotonically with an increase in the cladding hole radius. The larger pitch value, the larger hole radius required to achieve a low loss level.

What is the reason for such a striking difference between the behavior of waveguide losses in ASBGFs and HFs? In order to answer this question, it is necessary to examine the behavior of the spin and orbital parts of the Poynting vector of the fundamental core mode in both types of the fibers.

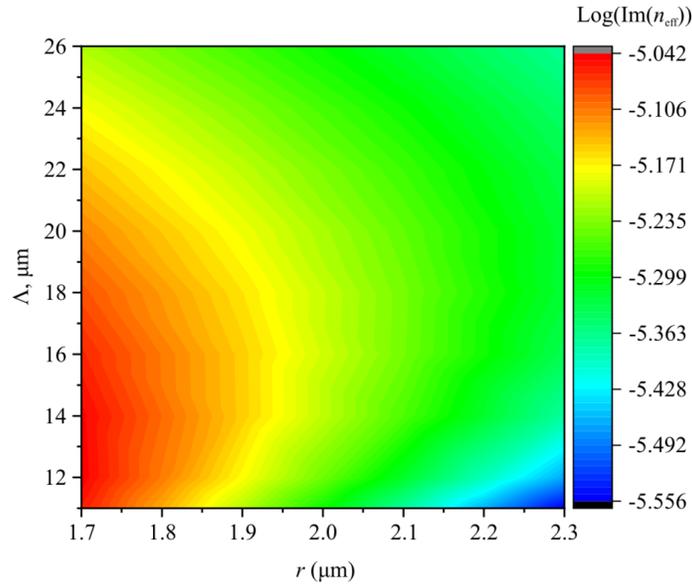

Fig. 5. Dependence of waveguide losses of the fundamental core mode of an HF with one row of the cladding holes on values of the cladding hole radius and pitch; the refractive index of the glass matrix is 1.45. The calculation was carried out at a wavelength of 1 μm for circular polarization. Cross – section of the fiber is shown in Fig. 3.

Let us consider behavior of spin and orbital parts of the transverse component of the Poynting vector of the fundamental core mode of an ASBGF with a cladding rode diameter of $r = 2$ μm and $\Lambda = 12$ μm. This is a lower part of the 'horseshoe' displaying the low loss area (Fig. 4). In this case, the waveguide loss of the ASBGF is about 1 dB/m. The spin and orbital parts of the transverse component of the Poynting vector are shown in Fig. 6 and 7 for linear and circular polarization of the fundamental core mode.

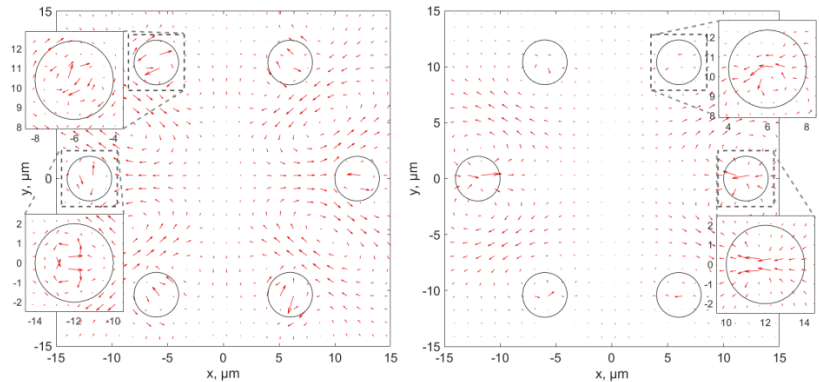

Fig. 6. Spin (left) and orbital (right) parts of the transverse component of the Poynting vector of the fundamental core mode of an ASBGF with parameters given in the text for linear polarization. The calculation was made at a wavelength of 1 μm.

Fig. 6 shows that the spin part of the Poynting vector in the case of linear polarization of the fundamental core mode has vortex features in the cladding rods. The orbital part of the

Poynting vector also has vortex features in the cladding rods, while in the rods located on the Y = 0 axis there is a strong movement of the fundamental core mode energy to the center of the fiber.

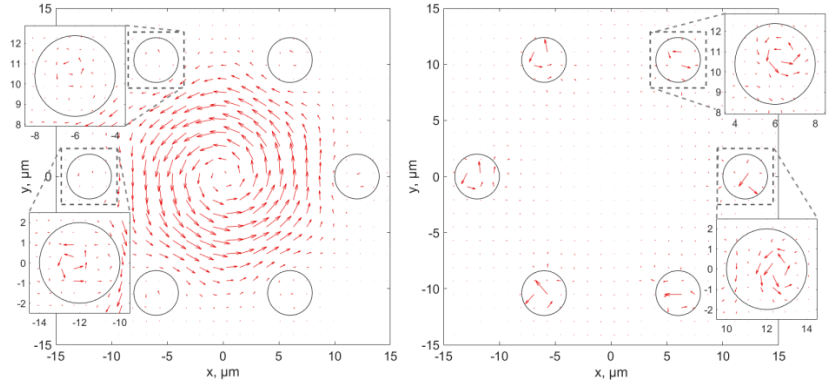

Fig. 7. Spin (left) and orbital (right) parts of the transverse component of the Poynting vector of the fundamental core mode of the ASBGF with parameters given in the text for circular polarization. The calculation was made at wavelength of 1 μm.

For circular polarization of the fundamental core mode, the rotation is observed for both the spin and orbital parts of the transverse component of the Poynting vector. For the spin part of the transverse component of the Poynting vector, the rotation is carried out as in the case of a dielectric all - solid pipe (Fig. 2(left)), around a singularity lying on the axis of the fiber. For the orbital part of the transverse component of the Poynting vector, the rotation is carried out mainly around the singularities lying inside the cladding rods.

To determine which part of the flux in the case of ASBGF mostly contributes to the waveguide losses, the total flow of the spin and orbital parts of the transverse component of the Poynting vector of the fundamental core mode was calculated through the boundary of a 36 μm - side square at geometric parameters of an ASBGF corresponding to the loss minimum (described above) (Fig. 4). The transverse flux was normalized so that the longitudinal flux which was equal to 1 W. Then, corresponding wavelength dependences were calculated and compared with the fiber loss in this spectral range (Fig. 8). It can be seen from Fig. 8 that waveguide losses are mainly determined by the orbital part of the transverse component of the Poynting vector of the fundamental core mode as in the case of an all - solid dielectric pipe. The spin part of the transverse component of the Poynting vector can be considered as a small correction to the waveguide losses of the fiber.

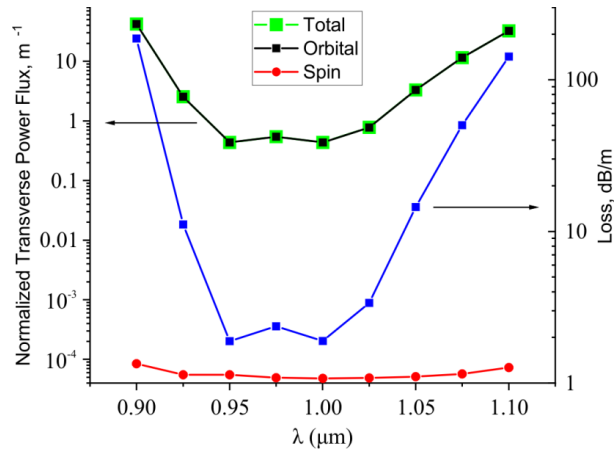

Fig. 8. Waveguide loss dependence of the spin and orbital parts of the transverse component of the Poynting vector of the fundamental core mode on a wavelength for ASBGF described in the text (circular polarization) in the spectral region near the loss minimum.

The distributions of the spin and orbital parts of the transverse component of the Poynting vector of the fundamental core mode for an HF with the same geometric parameters as for the ASBGF (waveguide loss minimum) in the case of linear polarization is shown in Fig. 9.

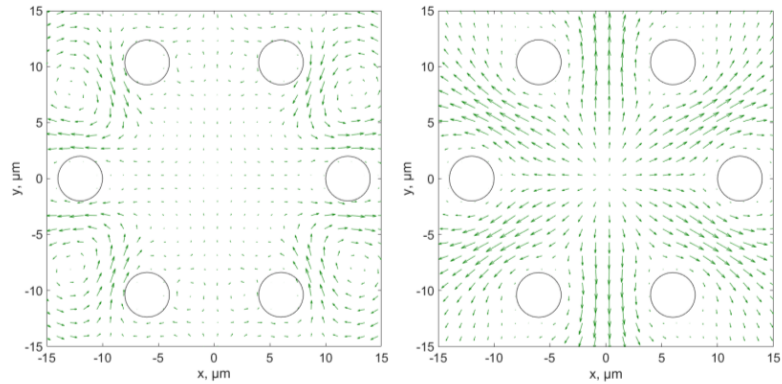

Fig. 9. Spin (left) and orbital (right) parts of the transverse component of the Poynting vector of the fundamental core mode of an HF with the same geometric parameters as for the ASBGF at a loss minimum (Fig. 4) in the case of linear polarization. The calculation was carried out at a wavelength of 1 μm.

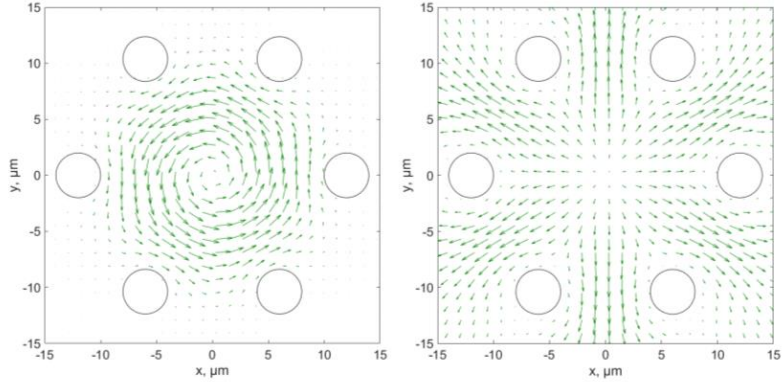

Fig. 10. Spin (left) and orbital (right) parts of the transverse component of the Poynting vector of the fundamental core mode of an HF with the same geometric parameters as for ASBGF in the case of circular polarization. The calculation was made at a wavelength of 1 μm.

For circular polarization, the distribution of the spin and orbital parts of the Poynting vector are shown in Fig. 10. It can be seen from Fig. 9 and 10 that the spin part has a vortex configuration, while the orbital parts behave similarly for both polarizations and determine the mode energy outflow in the space between the cladding holes. This behavior can explain a monotonic decrease in waveguide losses in HFs (Fig. 5) when only an increase in the cladding hole radius result in a decrease in the value of the orbital part of the transverse component of the Poynting vector flowing between the holes.

Also, an increase in the number of the cladding holes in the HF cladding will lead to a decrease in waveguide losses but the pattern of the fundamental core mode energy leakage will not change qualitatively. In the case of an ASBGF, new vortex motions of the fundamental core mode energy will appear in the cladding rods, directing the energy to the center of the fiber.

## 4. Conclusions

Based on the results obtained in this paper, it can be argued that singular points of the transverse profile of the transverse component of the Poynting vector are fundamental characteristics of the leaky waveguide core modes. Sets of singular points can form a 'singular skeleton' of the transverse component of the Poynting vector of ASBGF core modes. The spin part of the Poynting vector of the fundamental core mode of ASBGFs and HFs mainly contributes to its axial component of the core mode energy flux and only small part determines the core mode energy movement in the transverse direction. The orbital part contributes to all components of the Poynting vector, in particular, it manly determines the core mode energy flux in the transverse direction and, correspondingly, the waveguide loss of ASBGFs and HFs. A further study of the 'singular skeleton' properties of orbital part of the transverse component of the Poynting vector can serve as an efficient tool for establishing the relationship between the cross − section structures of SC MOFs and the position of singular points at which minimal losses are obtained for leaky mode in a given spectral range.

## Acknowledgements

The work was financially supported by RSF grant N 22 – 22 – 00575.

## Disclosures

The authors declare no conflicts of interest.